\documentclass{ws-mpla}
\usepackage[super]{cite}
\usepackage{graphicx}
\usepackage{mathtools}
\usepackage{hyperref}
\usepackage{float}
\begin{document}

\markboth{Parbati
Sahoo, S. Bhattacharjee, S. K. Tripathy, P.K. Sahoo}
{Matter bounce in $f(R,T)$ gravity}

%%%%%%%%%%%%%%%%%%%%% Publisher's Area please ignore %%%%%%%%%%%%%%
\catchline{}{}{}{} 
%%%%%%%%%%%%%%%%%%%%%%%%%%%%%%%%%%%%%%%%%%%%%%%%%%%%%%%%%%%%%%%%%%%

\title{Bouncing scenario in $f(R,T)$
gravity}

\author{\footnotesize Parbati
Sahoo}
\address{Department of Mathematics,\\ National Institute of Technology Calicut,\\ Kozhikode-673601, Kerala,
India,\\  Email:  sahooparbati1990@gmail.com}

\author{S. Bhattacharjee}

\address{Department of Astronomy,\\ Osmania University,\\ Hyderabad-500007,
India,\\  Email: snehasish.bhattacharjee.666@gmail.com}

\author{S. K. Tripathy}
\address{Department of Physics,\\ Indira Gandhi Institute of Technology, Sarang,\\ Dhenkanal, Odisha 759146, India,\\ Email: tripathy\_sunil@rediffmail.com}

\author{P.K. Sahoo}

\address{Department of Mathematics,\\ Birla Institute of
Technology and Science-Pilani, \\ Hyderabad Campus, Hyderabad-500078,
India\\
pksahoo@hyderabad.bits-pilani.ac.in}

\maketitle

\pub{Received (30 August 2019)}{Accepted (18 December 2019)}

\begin{abstract}
The present manuscript presents modeling of matter bounce in the framework of $f(R,T)$ gravity where $f(R,T) = R + 2 \lambda T$. We start by defining a parametrization of scale factor which is non-vanishing. The geometrical parameters such as the Hubble parameter and deceleration parameter are derived, from which expressions of pressure, density and Equation of State (EoS) parameter and a qualitative understanding of the initial conditions of the universe at the bounce are ascertained. We found that the initial conditions of the universe are finite owing to the non-vanishing nature of the scale factor thus eliminates the initial singularity problem. Furthermore, we show the violation of energy conditions near the bouncing region and analyzed the stability of our model with respect to linear homogeneous perturbations in Friedmann-Lema\^tre-Robertson-Walker (FLRW) spacetime. We found that our model and hence matter bounce scenarios in general are highly unstable at the bounce in the framework of $f(R,T)$ gravity but the perturbations decay out rapidly away from the bounce safeguarding its stability at late times.

\end{abstract}

\ccode{PACS Nos.: 04.50.kd}

\section{Introduction}
It is well known that Einstein’s general theory of gravity (GR) \cite{gr} is one of the most elegant theory in all of science. The theory revolutionizes the way we think of gravity. We now know that gravity is not some force emanating from objects as Newton first postulated, rather some distortion in the fabric of space-time caused by the distribution of matter. GR essentially states that an accelerated frame of reference is equivalent to a gravitational field, thus is an extension of special theory of relativity \cite{sr} which could only work for uniform motion.

Some of the observational evidences of GR include distorted images of astrophysical objects caused by gravitational lensing, existence of supermassive black hole (Sagittarius A) at the center of milky way inferred through Doppler imaging of highly elliptical orbits and superfast motions of its nearby stars \cite{motion}, recent images of supermassive black hole at the heart of $M87$ \cite{eht}, gravitational redshift of electromagnetic waves and detection of gravitational waves from the collisions of compact stars by LIGO \cite{ligo}. 

Though the theory stood the test of time and has diverse applications in physical cosmology, it cannot explain the biggest problem in physical cosmology, i.e the current acceleration of the universe. The acceleration of the universe at the present epoch cannot be explained without invoking new forms of matter-energy fields\cite{prd}. This exotic entity is termed Dark energy (DE). Theories emerged to decode this enigma by proposing various candidates for DE such as quintessence, spintessence, tachyons, f-essence, k-essence,  phantom, Chaplygin gas \cite{candidate}.

Despite these convincing models none of these hypothetical candidates have been directly observed nor produced in the terrestrial laboratory. Thus researchers were motivated to drop this idea of DE and started modifying the geometrical sector of the Einstein's field equations. By rearranging the Einstein - Hilbert action, new modified theories of gravity emerged capable of mimicking the late time acceleration of the universe. Some of these modified theories are $f(\mathcal{T})$ gravity \cite{t}, where $\mathcal{T}$ is the torsion scalar, $f(R)$ gravity \cite{r1,r2,r3,r4,r5}, where $R$ is the Ricci scalar, $f(R,T)$ gravity \cite{harko/2011} where $R$ is the Ricci scalar, $T$ is the trace of the stress energy-momentum tensor, $f(G)$ gravity \cite{g} where G is the Gauss-Bonnet invariant, etc. 

$f(R,T)$ gravity models are frequently studied in the literature due to its robustness in solving many cosmological as well as astrophysical problems \cite{pk}. In $f(R,T)$ modified gravity the matter Lagrangian $L_{m}$ is varied with respect to the metric which is represented by the presence of a source term. Expression of this source term is obtained as a function of $T$, hence different choices of $T$ would generate different set of field equations. In this model the covariant divergence of stress energy momentum tensor does not vanish, hence the motion of classical particles does not follow geodesics resulting in an extra acceleration which suffices the late time acceleration of the universe without adopting to DE but the law of energy momentum conservation is sacrificed.

In the paradigm of big bang cosmology, our universe emerged out of a singularity, is finite in time and space and is around $13.7$ billion years young. Though the historic discovery of CMB by A.Penzias and R.Wilson \cite{cmb} supports big-bang model, it has a number of shortcomings such as flatness problem, horizon problem, entropy problem, transplanckian problem, singularity problem and original structure problem. Alan Guth proposed the theory of inflation in which  the universe is believed to have underwent exponential expansion for a very short period of time ($10^{-30}$ sec) shortly after the big bang \cite{inflation, instar}. The inflationary scenario can mimic the observations of CMB due to the flexibility of its parameters \cite{flexible}. Though the inflationary scenario could be able to address many of the above mentioned problems, the singularity problem still remain unanswered. Thus instead of inflationary models, we focus on alternative scenarios of formation and evolution of the universe, namely the cyclic universe which states that our universe transpired from a prior contracting phase and is destined to undergo an expanding phase without suffering from any singularity, or in other words it undergoes a bouncing phase. Many authors \cite{c1,c2,c3,c4,c5,rb} have studied diverse phenomenological features of the bouncing scenario such as a single scalar field matter containing a kinetic and potential term, a contracting universe consisting of radiation, bounce model with dark matter and dark energy, observational bouncing cosmologies with Planck and BICEP2 data and the characteristics of bouncing cosmology as alternative theories to the inflation which are in harmony with observations. Bamba et al., \cite{b1,b2,b3,b4} have studied bouncing cosmologies in $f(R)$ gravity, $f(\mathcal{T})$ gravity and in $f(G)$ gravity and examined the dynamical stability of the solutions.  de la Cruz-Dombriz et al. \cite{c6} reported bouncing cosmology model in teleparallel gravity. Cai et al. \cite{c7} have studied bouncing models in $f(\mathcal{T})$ gravity. Tripathy et al. have studied some bouncing models in $f(R,T)$ gravity theory and obtained that, the matter-geometry coupling constant appearing in the modified geometrical action has a substantial affect on the cosmic dynamics near bounce  \cite{SKT19}.

The paper is organized as follows: In Section II we present an overview of $f(R,T)$ gravity. In Section III we introduce the concept of matter bounce. We define a parametrization of scale factor and study the detailed dynamical evolution of an universe undergoing a non-singular bounce. In Section IV we study the violation of energy conditions. In Section V we analyze the stability of our model with respect to linear homogeneous perturbations in FLRW spacetime. Finally, in Section VI we present a summary and conclude the work.
\section{Field equations and Solutions}
For the $f(R,T)$ gravity formalism, the geometrically modified action with matter is given by 

\begin{equation}\label{e1}
S=\int  d^{4}x \sqrt{-g}\left[\frac{1}{2} f(R,T)+\mathcal{L}_{m} \right] .  
\end{equation}

We set $8\pi G=c=1$; where $G$ and $c$ are Newtonian gravitational constant and speed of light. $\mathcal{L}_m$ is the matter Lagrangian density related to stress-energy tensor as
\begin{equation}\label{e2}
T_{ij}= - \frac{2}{\sqrt{-g}} \frac{\delta(\sqrt{-g}\mathcal{L}_{m})}{\delta g^{ij}}.
\end{equation}

By varying the action $S$ given in (\ref{e1}) with respect to metric $g_{ij}$ provides the $f(R,T)$ field equations \cite{harko/2011}
\begin{multline}\label{e4}
F(R,T)\left(R_{ij}-\frac{1}{3} Rg_{ij}\right) + \frac{1}{6}f(R,T)g_{ij} \\= \left(T_{ij}-\frac{1}{3}Tg_{ij}\right)-\mathcal{F}(R,T)\left(T_{ij} -\frac{1}{3}Tg_{ij}\right)\\-\mathcal{F}(R,T)\left(\theta_{ij}-\frac{1}{3}\theta g_{ij}\right)+\nabla_i\nabla_jF(R,T).
\end{multline}

Here, the notations are $F(R,T)=\partial f(R,T)/\partial R$ and $\mathcal{F}(R,T)=\partial f(R,T)/\partial T$ respectively and
\begin{equation}\label{e5}
\theta_{ij}=g^{ij}\frac{\delta T_{ij}}{\delta g^{ij}}.
\end{equation}
The matter Lagrangian is considered as $\mathcal{L}_m=-p$, where $p$ is the pressure. Hence, equation (\ref{e5}) can be written as \cite{harko/2011}
\begin{equation}\label{e6}
\theta_{ij}=-2T_{ij}-p g_{ij}.
\end{equation}
%The $f(R,T)$ gravity field equations (\ref{e4}) for the linear %choice of $f(R)$ i.e. $f(R,T)=R+ 2 \lambda T$ with (\ref{e6}) %takes the form
 
%\begin{equation}\label{7}
%R_{ij}-\frac{1}{2}Rg_{ij}= T_{ij}+2f'(T)T_{ij}+[2p f'(T)%%%%+f(T)]g_{ij}
%\end{equation}
%Assuming $f(T)=2\lambda T$, where $\lambda$ is constant, the %above equation can be written as
%\begin{equation}\label{8}
%R_{ij}-\frac{1}{2}Rg_{ij}=(1+2\lambda) T_{ij}+(2p+T)\lambda %g_{ij}
%\end{equation}

The spatially homogeneous and flat FLRW metric is given as
\begin{equation} \label{1}
ds^{2}=dt^{2}-a^{2}(dx^{2}+dy^{2}+dz^{2}),
\end{equation}
where $a(t)$ is known as cosmic scale factor.\\
The energy momentum tensor for perfect fluid matter is taken as in this form
\begin{equation}\label{2}
T_{ij}=(\rho+p)u_{i}u_{j}-pg_{ij}.
\end{equation}\\
Thus the modified Friedmann equations for for the linear choice of $f(R,T)$ i.e. $f(R,T)=R+ 2 \lambda T$ for a perfect fluid distribution in a FLRW background takes the form

\begin{eqnarray}
3H^2=(1+3\lambda)\rho-\lambda p, \label{eqn1}\\
2\dot{H}+3H^2=-(1+3\lambda)p+\lambda \rho, \label{eqn2}
\end{eqnarray}
where dots represented as the derivatives with respect to time $t$.\\
Using equations (\ref{eqn1}) and (\ref{eqn2}) we obtain the energy density $\rho$, pressure $p$ and EoS parameter $\omega=p/\rho$ respectively as
\begin{equation}\label{eqn3}
\rho=\frac{-2 \dot{H} \lambda +3(1+2\lambda)H^2}{(1+3 \lambda)^2- \lambda^2},
\end{equation}
\begin{equation}\label{eqn4}
p=-\frac{3(1+2\lambda)H^2+2(1+3\lambda)\dot{H}}{(1+3 \lambda)^2- \lambda^2},
\end{equation}
\begin{equation}\label{eqn5}
\omega=-\frac{3(1+2\lambda)H^2+2(1+3\lambda)\dot{H}}{-2 \dot{H} \lambda +3(1+2\lambda)H^2}.
\end{equation}
The dynamical behavior of the physical parameters like energy density, pressure and EoS parameter depends on the behavior of Hubble parameter and $\lambda$. The EoS parameter reduces to GR for a vanishing $\lambda$.

\section{Matter Bounce Scenario}
Matter bounce scenarios are a set of cosmological models comprising an initial contracted matter-dominated state coupled with a non-singular bounce \cite{h31}. Such bouncing models are reported to provide an exiting alternative to inflation by reproducing the observed spectrum of cosmological fluctuations \cite{h19,h19a,h19b,h32,h33}. In these models, the Strong Energy Condition is violated  near bouncing epoch by introducing new forms of matter in the framework of GR. Therefore, in order to investigate bouncing dynamics keeping the matter sector unchanged, one has to go beyond GR. Matter bounce have been studied employing quintom matter \cite{h34}, Galileon fields \cite{h37}, Lee-Wick matter \cite{h35} and phantom field \cite{h38}.

Matter bounce scenarios can be described by the general expression
\begin{equation}
a(t)=a_{0}\left(M \rho_{cr} t^{2} + Q \right) ^{Z},
\end{equation}
where $M$, $Q$, $Z$ are constants and $\rho_{cr} = \frac{3 H_{0}^{2}}{8 \pi G}= 1.88 h^{2} \times 10^{-29}g cm^{-3}$ is the critical density of the universe \cite{michael}. $M = 2/3, 3/2, 3/4$ or $4/3$, $Q = 1$ and $Z = 1/3$ or $1/4$ \cite{h41,h42,h44,h86}. 
 
\subsection{Scale factor}
Here we work with a scale factor of the form \cite{c6}
\begin{equation}\label{m1eqn1}
a(t)=a_{0}\left[ \frac{3}{2}\rho_{cr} t^{2} +1\right] ^{1/3},
\end{equation}
where $a_{0}$ is the value of the scale factor at the transfer point (bouncing epoch). It can also be verified that $\dot{a}<0$ for $t<0$ and $\dot{a}>0$ for $t>0$. Thus the universe transits from a prior contracting phase ($t<0$) to a later expanding phase ($t>0$).

The non-vanishing scale factor at $t=0$ ensures a null value of Hubble parameter at the transfer point. For a successful bouncing scenario, the Hubble parameter must be negative before the bounce and positive after the bounce with $\dot{H}  = 4 \pi G \rho (\omega + 1)> 0$ in the vicinity of the bouncing epoch. Note that to satisfy this inequality, $\omega < -1$ and enters the phantom region around the transfer point and violate the Null Energy condition (NEC). 
\begin{figure}[H]
\centering
  \centering
  \includegraphics[width=85 mm]{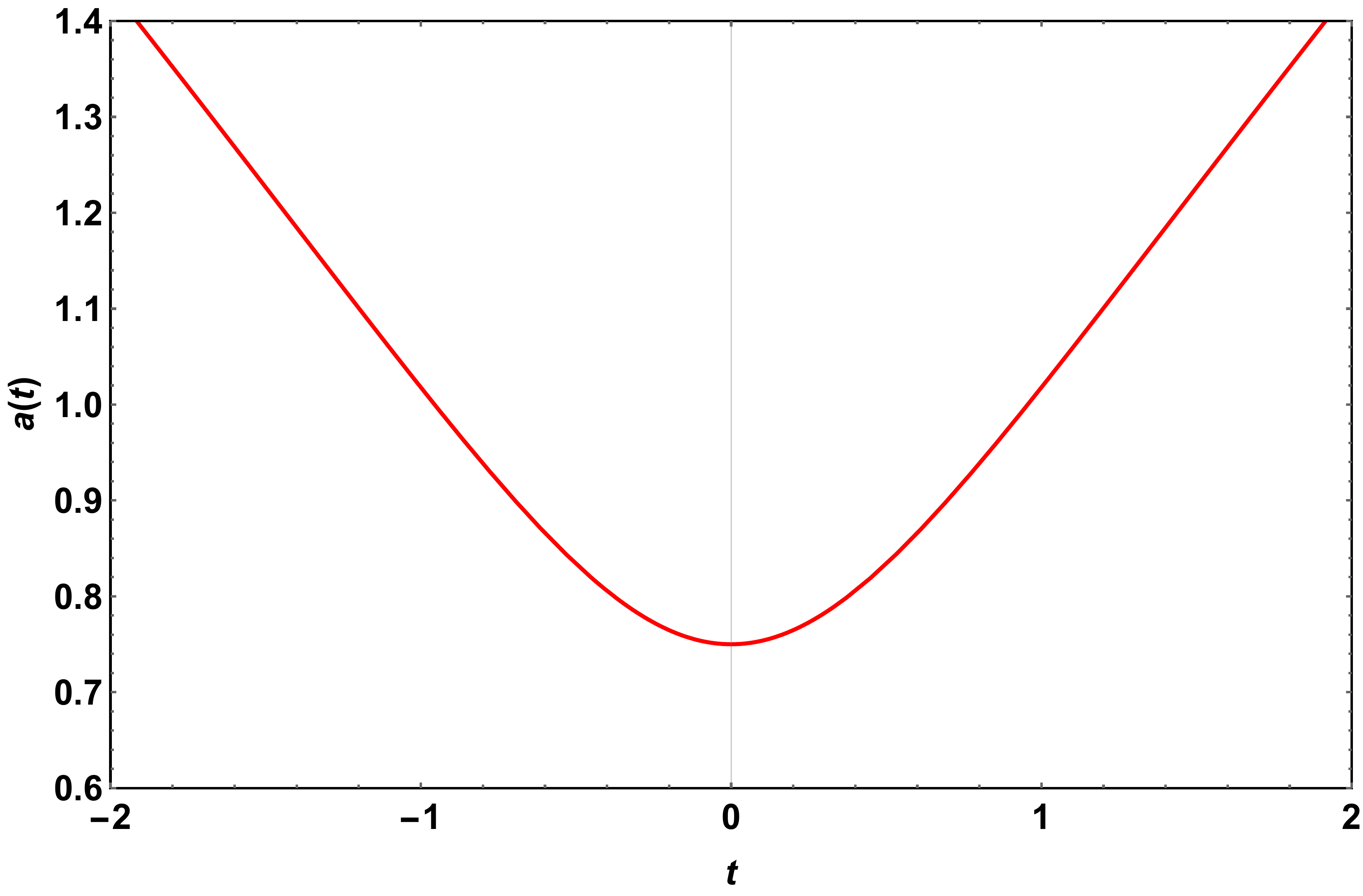}
  \caption{Time evolution of scale factor with $a_{0}= 0.75$.}
  \label{m1fig1}
\end{figure}
Since we are interested in understanding the initial conditions at the bouncing epoch and time evolution of cosmological parameters, we set $\rho_{cr} = 1$ as it is a constant and time is represented in units of $\sqrt{\frac{1}{\rho_{cr}}}$.
\subsection{Hubble parameter and Deceleration parameter}
The Hubble parameter for \eqref{m1eqn1} reads  
\begin{equation}\label{m1eqn2}
H=\frac{2t \rho_{cr}}{2+3 t^{2} \rho_{cr}}.
\end{equation}
The maximum value of H is achieved when $t = t_{max} = \pm \sqrt{\frac{2}{3 \rho_{cr}}}$, with $H = \pm \sqrt{\frac{\rho_{cr}}{6}}$. The maximal value of torsion scalar is $T = T_{max} = 6 H_{max}^{2}= \rho_{cr}$. The free parameter of \eqref{m1eqn1} can be constrained with observational data by defining the current time $t = t_{0}$ where $a = a_{0}= 1$ and reads
\begin{equation}
\frac{2}{3 \rho_{cr}}\left(\frac{1}{a_{0}^{3}} - 1 \right) =t_{0}^{2}.
\end{equation}
Since the critical density is a positive quantity, the equality holds as long as $0 < a_{0} < 1$.

The deceleration parameter ($q = \frac{-\ddot{a} a}{\dot{a}^{2}}$) reads  
 \begin{equation}\label{m1eqn3}
 q=\frac{1}{2}-\frac{1}{\rho_{cr} t^2}.
\end{equation}
\begin{figure}[H]
  \centering
  \includegraphics[width=85 mm]{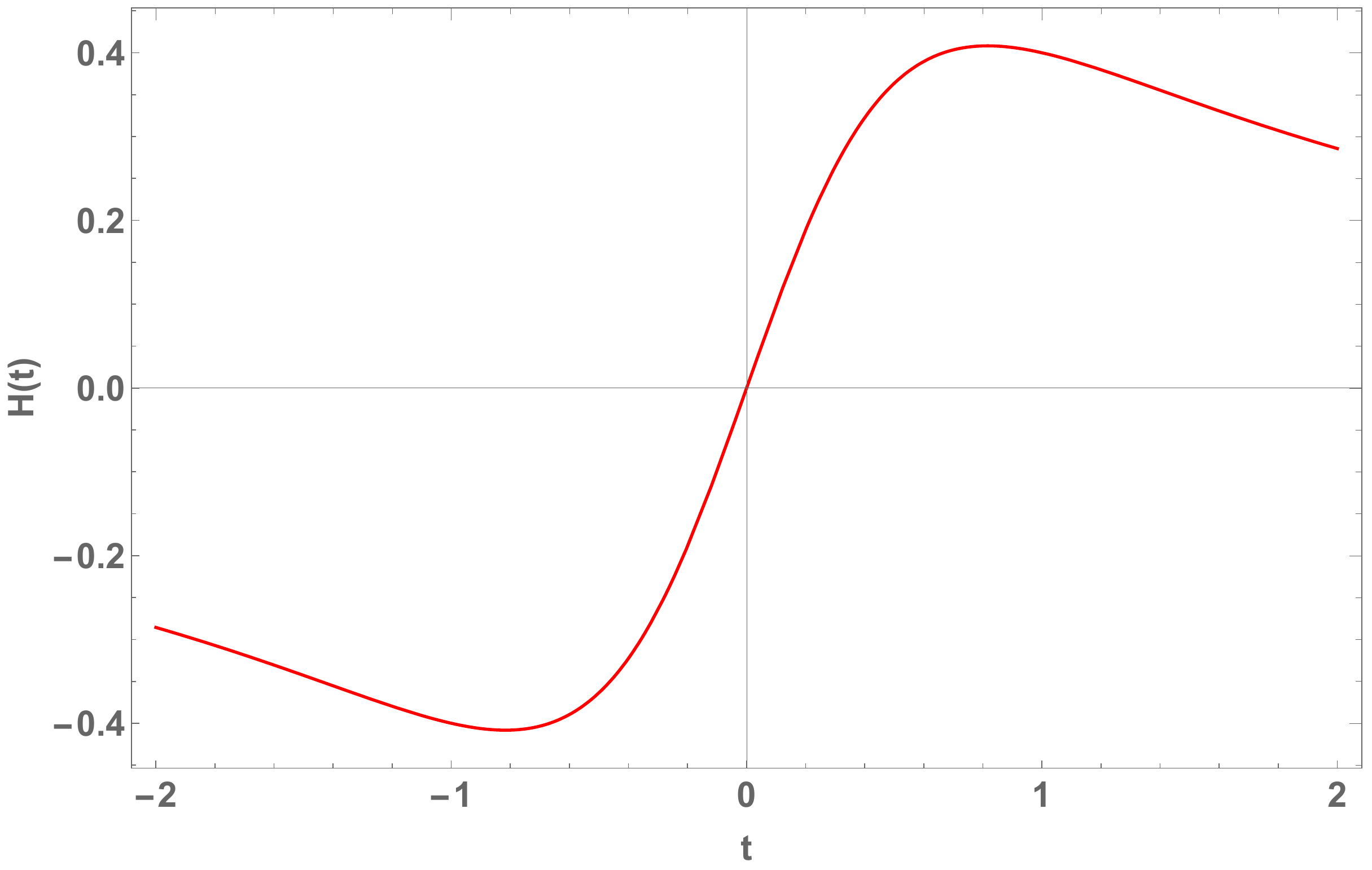}
  \caption{Time evolution of Hubble parameter.}
  \label{m1fig2}
\end{figure}
%\begin{figure}[H]
 % \centering
  %\includegraphics[width=85 mm]{q-critical.pdf}
  %\caption{Evolution of deceleration parameter against $t$.}
  %\label{m1fig3}
%\end{figure}
\subsection{EoS parameter and Constraining $\lambda$}
The EoS parameter is a very useful quantity in understanding the viability of a bouncing scenario. Substituting \eqref{m1eqn2} in \eqref{eqn5}, EoS parameter reads
\begin{equation}\label{m1eqn6}
\omega =\frac{3 \lambda  \left(\rho_{cr} t^2-2\right)-2}{3 (3 \lambda +1) \rho_{cr} t^2-2 \lambda }.
\end{equation}
Note that at the bouncing epoch ($t=0$) for $\lambda = 0$, $\omega\rightarrow \infty $. Therefore, to achieve a successful matter bounce without employing exotic matter energy fields, one needs to introduce new physics near the bouncing region, which in this case is a modification of GR, namely $f(R,T)$ gravity.

The model parameter $\lambda$ can be constrained from the condition that $\omega < -1$ near the bouncing epoch. Substituting $t=0$ in \eqref{m1eqn6} yields
\begin{equation}\label{100}
\omega \bigg|_{t = 0} = \frac{-6 \lambda - 2}{- 2 \lambda}.
\end{equation}
For \eqref{100} to be $< -1$, $\lambda$ has to follow the restriction $0 > \lambda > -1/4$.
%\begin{equation}\label{m1eqn7}
%\rho+p=\frac{4\rho_{cr} \left(3 \rho_{cr} t^2-2\right)}{(2 %\lambda +1) \left(3 \rho_{cr} t^2+2\right)^2}
%\end{equation}
\begin{figure}[H]
  \centering
  \includegraphics[width=80 mm]{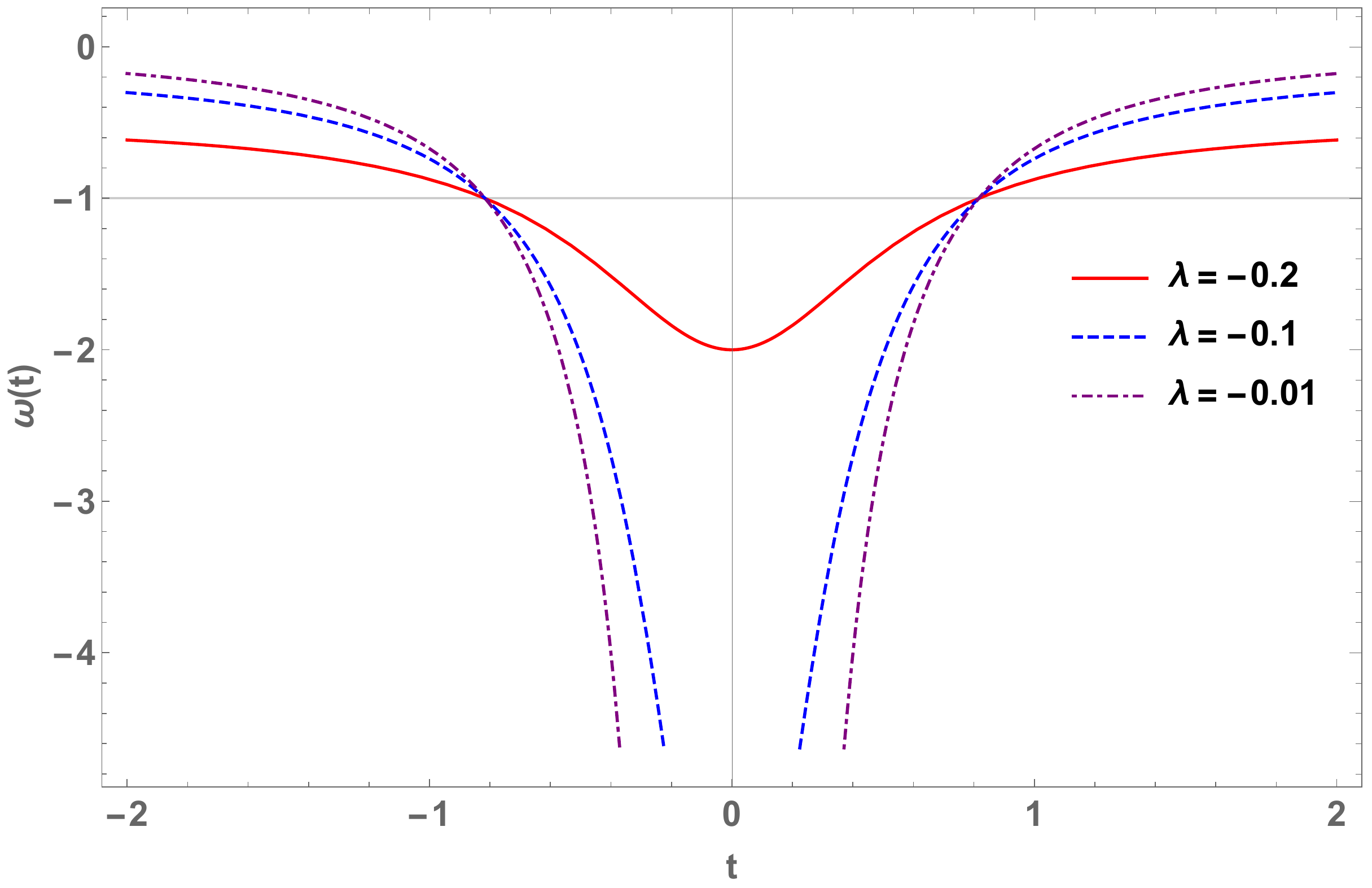}
  \caption{Time evolution of EoS parameter.}
  \label{m1fig6}
\end{figure}
An expression of time $t$ when $\omega$ crosses the phantom divide line can be obtained by equating the R.H.S of \eqref{m1eqn6} to be equal to -1, which turns out to be
\begin{equation}
t \bigg |_{\omega = -1} = \pm \sqrt{\frac{2}{3\rho_{cr}}}
\end{equation}
\subsection{Initial Conditions}
In the big bang cosmology, the initial conditions of the universe is unknown as scale factor vanishes making the density infinite. Since in non-singular bouncing cosmology, the scale factor does not vanish, the initial conditions such as pressure and density can be well understood. This can be done by assuming an prior ansatz of scale factor and plugging it in the Friedmann equations and extrapolating it backwards to find the necessary conditions for a successful bouncing scenario. 

Substituting \eqref{m1eqn1} in \eqref{eqn3} and \eqref{eqn4}, the expressions of pressure and density reads
\begin{equation}\label{m1eqn4}
\rho=\frac{12\rho_{cr}^2 (3 \lambda +1) t^2-8 \lambda \rho_{cr} }{(4 \lambda +1)(2\lambda +1) \left(3 \rho_{cr} t^2+2\right)^2},
\end{equation}

\begin{equation}\label{m1eqn5}
p=\frac{12 \rho_{cr}^2 \lambda t^2-24\rho_{cr} \lambda-8\rho_{cr}}{(4 \lambda +1)(2\lambda +1) \left(3 \rho_{cr} t^2+2\right)^2}.
\end{equation} 
\begin{figure}[H]
\centering
   \includegraphics[width=75 mm]{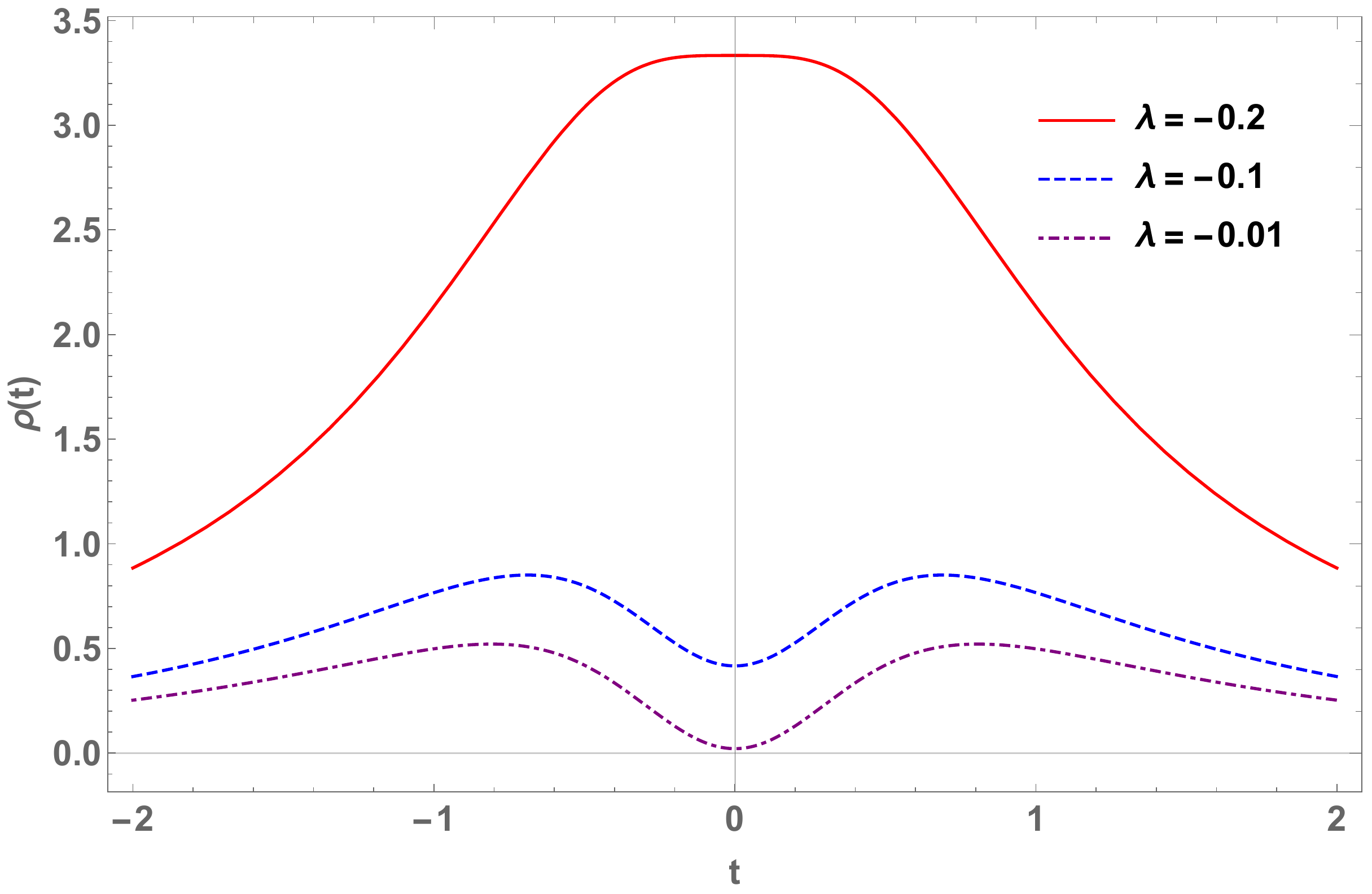}
  \caption{Time evolution of energy density.}
  \label{m1fig4}
\end{figure}
\begin{figure}[H]
  \centering
  \includegraphics[width=75 mm]{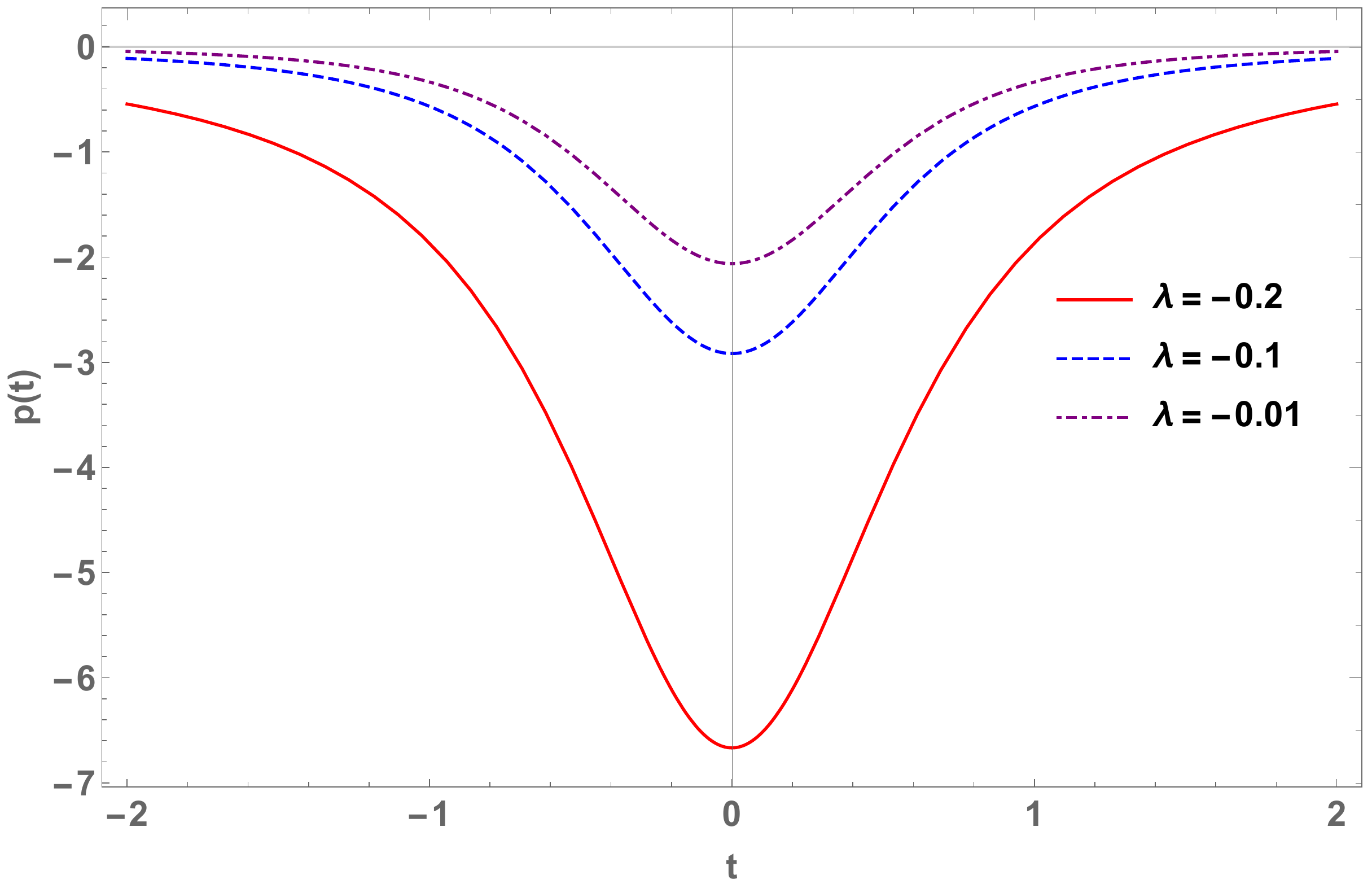}
  \caption{Time evolution of pressure.}
  \label{m1fig5}
\end{figure}
An increase in $\lambda$ decreases the energy density and pressure near the bouncing territory. However, away from the bouncing territory the profiles start to converge.

Substituting $t=0$ in \eqref{m1eqn4} and \eqref{m1eqn5}, we obtain the values of pressure and density at the bouncing epoch and reads
\begin{equation}\label{99}
p =- \rho_{cr} \left( \frac{1 + 3 \lambda}{(1 + 4 \lambda)(2\lambda + 1)}\right), 
\end{equation}
\begin{equation}\label{98}
\rho = \frac{-2 \lambda \rho_{cr}}{(1 + 4 \lambda)(2\lambda + 1)}.
\end{equation}
Substituting $\lambda=0$ in \eqref{99} and \eqref{98}, we obtain 
\begin{equation}
p = - \rho_{cr}, \hspace{0.25in} \rho = 0.
\end{equation}
This once again demonstrates the need for GR modification to achieve a successful matter bounce.
\section{Violation of Energy Conditions}
Energy conditions (ECs) are a set of linear equations involving density and pressure which demonstrate that energy density cannot be negative and gravity is always attractive. ECs state that any linear combination of pressure and density cannot be negative \cite{sah60}. They are essestial in the studies of wormholes and thermodynamics of black holes and originate from the Raychaudhuri's equation \cite{sah61}. The ECs are expressed as 
\begin{enumerate}
\item Null Energy condition (NEC) $\Leftrightarrow \rho + p \geq 0$,
\item Strong Energy Condition (SEC) $\Leftrightarrow \rho + 3p \geq 0$,
\item Dominant Energy Condition (DEC) $\Leftrightarrow \rho > |p| \geq 0$,
\item Weak Energy Condition (WEC) $\Leftrightarrow \rho \geq 0$.
\end{enumerate} 
To achieve a successful non-singular bounce, the EoS parameter must cross the phantom divide ($\omega < -1$) and hence violate the NEC. The violation of energy conditions are shown in the following Figures \ref{m1fig7} , \ref{m1fig9}.  In the figures it is clear that, there is no singularity near the bouncing epoch. Rather, the energy conditions evolve symmetrically around the bouncing point. As apprehended for a matter bounce scenario, the energy conditions $\rho+p$ and $\rho+3p$ become negative near the bounce both in the positive and negative time zone and there is a clear indication of violation of energy condition leading to a situation where the model evolves in the phantom phase with $\omega <-1$.
\begin{figure}[H]
\centering
  \includegraphics[width=75 mm]{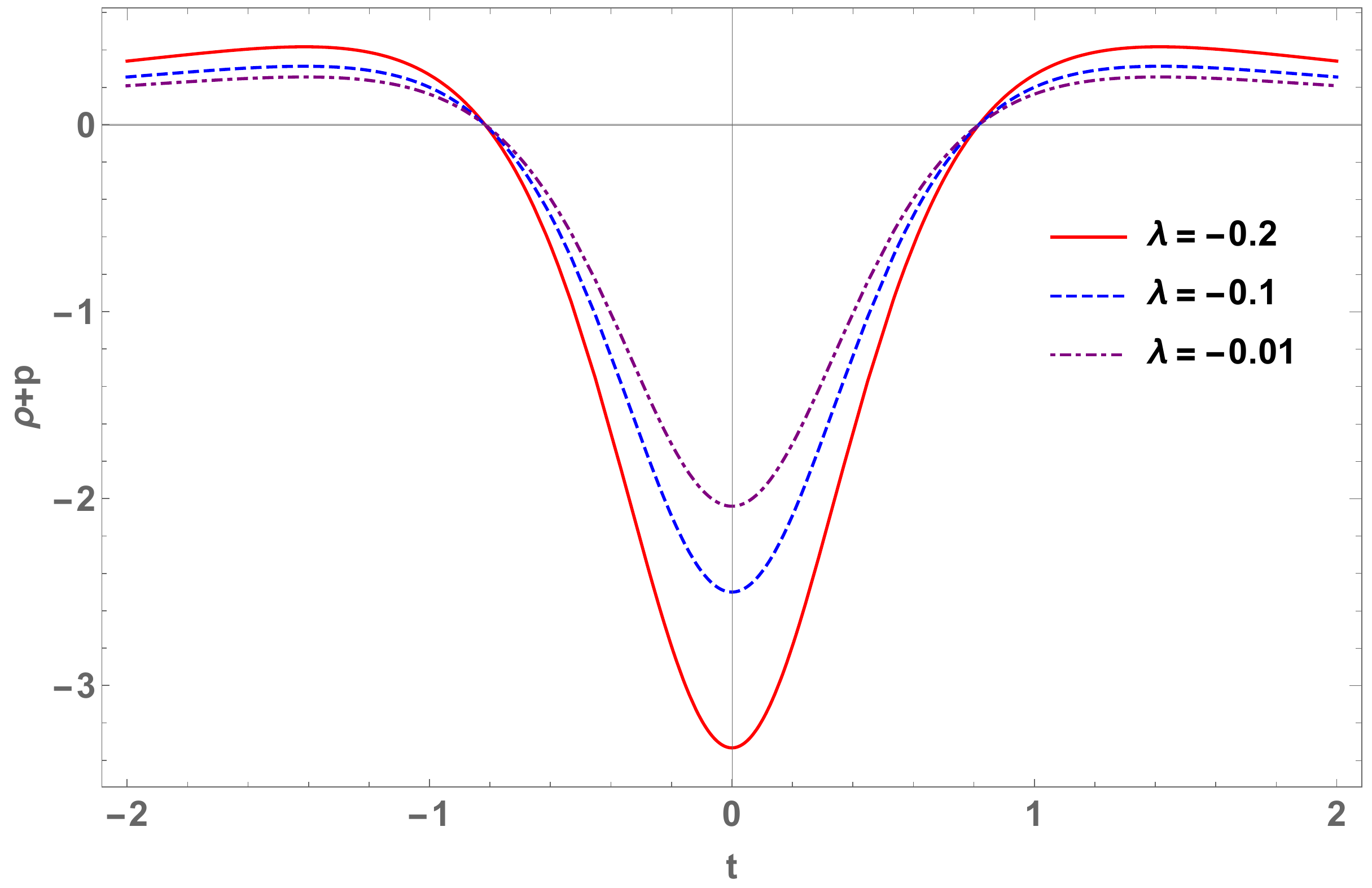}
  \caption{Violation of NEC at the bouncing region.}
  \label{m1fig7}
\end{figure}
%\begin{figure}[H]
  %\centering
  %\includegraphics[width=75 mm]{rho-p-critical1.pdf}
  %\caption{DEC.}
 % \label{m1fig8}
%\end{figure}
\begin{figure}[H]
  \centering
  \includegraphics[width=80 mm]{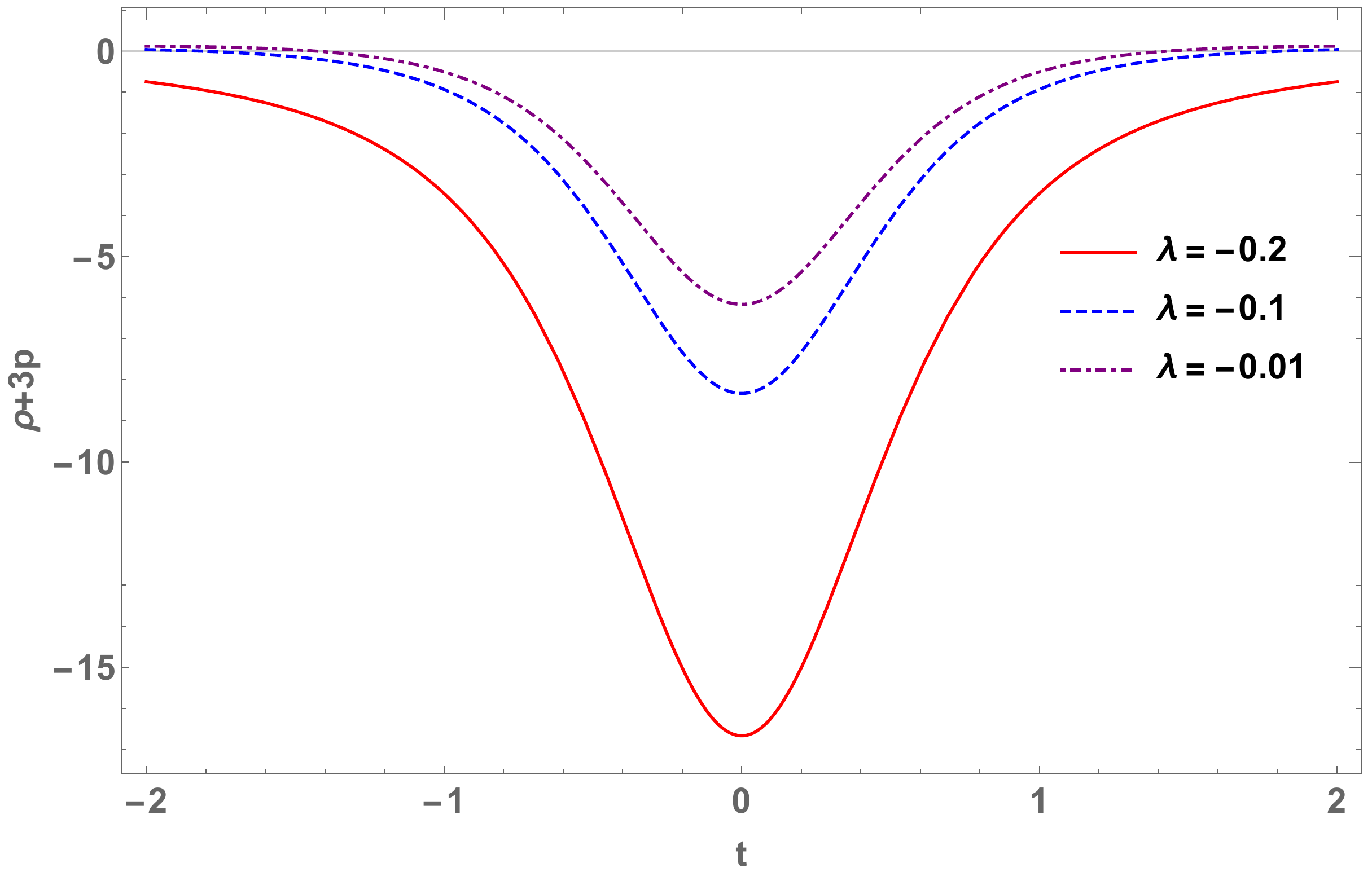}
  \caption{Violation of SEC at the bouncing region.} \label{m1fig9}
\end{figure}
\section{Stability Analysis}
We now seek to investigate the stability of \eqref{m1eqn1} under the influence of liner homogenous perturbations in the FLRW spacetime. Linear perturbations of Hubble parameter and energy density can be written as \cite{sah65}
\begin{equation}\label{97}
H(t) = H_{*}(t)(1 + \delta (t)),
\end{equation}
\begin{equation}
\rho (t) = \rho_{*}(1 + \delta_{n}(t)).
\end{equation}
where $\delta (t)$ \& $\delta_{n}(t)$ represent perturbation parameters. For $H(t) = H_{*}(t)$, the FLRW equations are satisfied. Expressing the matter density in terms of $H_{*}(t)$ as \cite{pk}
\begin{equation}
\rho_{*} = \frac{(6 \lambda + 3)H_{*}^{2} - 2 \lambda, \dot{)H_{*}}}{(1 + 3 \lambda)^{2} - \lambda^{2}}.
\end{equation}
The Friedmann equation and the trace equation in $f(R,T)$ gravity reads \cite{pk}
\begin{equation}\label{96}
\Omega^{2} = 3(2 \lambda (\rho + p) + \rho +  f(R,T)),
\end{equation}
\begin{equation}
R =  - 2 \lambda (\rho + p) - 4 f(R,T)-(\rho + 3 p),
\end{equation}
where $\Omega = 3 H$ denote the expansion scalar. The first order perturbation equation for a standard matter field is given as
\begin{equation}\label{95}
\dot{\delta}_{n} (t) + 3 H_{*} (t) \delta (t) = 0.
\end{equation}
Using Equations \eqref{97} -\eqref{96}, we obtain 
\begin{equation}\label{94}
T \delta_{n} (t) (1 + 3 \lambda) = 6 H_{*}^{2} \delta (t).
\end{equation}
Eliminating $\delta (t)$ from \eqref{95} and \eqref{94}, we obtain the first order perturbation equation as
\begin{equation}\label{93}
\dot{\delta}_{n} (t) + \frac{T}{2 H_{*}} ( 3 \lambda + 1 ) \delta_{n} (t) = 0.
\end{equation}
Integrating \eqref{93}, we obtain 
\begin{equation}\label{92}
\delta_{n} (t) = A \exp \left[-\left( \frac{ 3 \lambda + 1}{2}\right) \int \frac{T}{H_{*}} dt \right], 
\end{equation}
where $A$ is the integration constant. As a result, the evolutionary equation of perturbation is given by
\begin{equation}\label{91}
\delta (t) = \frac{( 3 \lambda + 1 ) A T}{6H_{*}^{2}} \exp \left[-\left( \frac{ 3 \lambda + 1}{2}\right) \int \frac{T}{H_{*}} dt \right].
\end{equation}
Note that at $t=0$, $H_{*} = 0$ and therefore \eqref{92} and \eqref{91} blow up making the model highly unstable near the bounce. However away from the bouncing region, the perturbations decay quickly ensuring stability at late times. We also note that \eqref{92} and \eqref{91} have valid solutions only for $t> 0$. For $t< 0$, the functions become imaginary.
\begin{figure}[H]
\centering
  \includegraphics[width=75 mm]{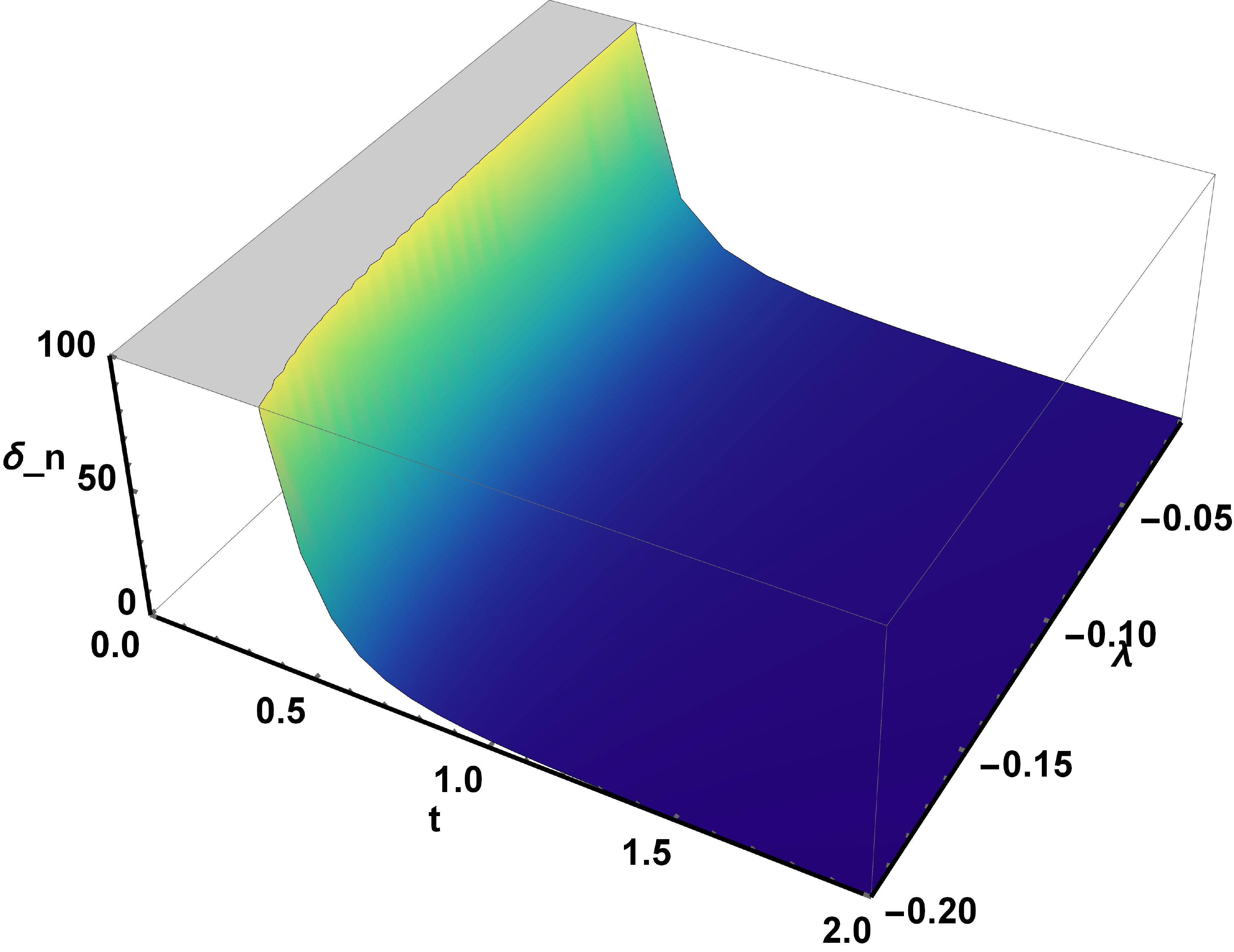}
  \caption{Evolution of $\delta_{n} (t)$.}
  \label{m1fig101}
\end{figure}
\begin{figure}[H]
\centering
  \includegraphics[width=75 mm]{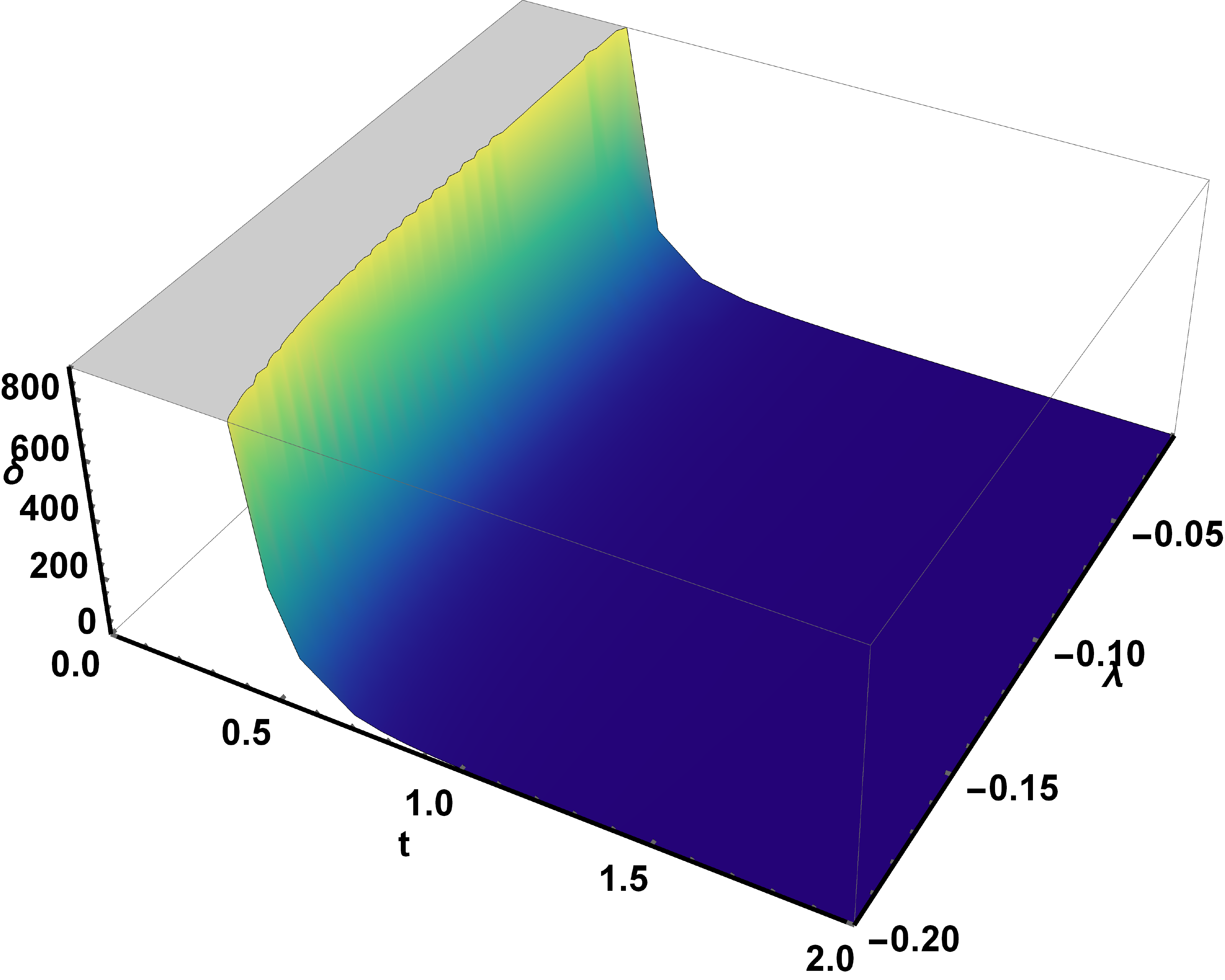}
  \caption{Evolution of $\delta (t)$.}
  \label{m1fig100}
\end{figure}

\section{Summary and Conclusion }
Matter bounce scenarios are a set of cosmological models comprising an initial contracted matter-dominated state coupled with a non-singular bounce \cite{h31} and provide a graceful alternative to inflation in the early universe. In this work we present a matter bounce scenario in the framework of $f(R,T)$ gravity where $f(R,T) = R + 2 \lambda T$. Our results are as follows:
\begin{itemize}
\item  By defining a parametrization of scale factor of the form $a(t)=a_{0}\left( \frac{3}{2}\rho_{cr} t^{2} +1\right) ^{1/3}$, the geometrical parameters such as the Hubble parameter and deceleration parameter are derived.
\item The EoS parameter $\omega$ is derived by substituting the expression of Hubble parameter \eqref{m1eqn2} into \eqref{eqn5} and the model parameter $\lambda$ is constrained from the condition that at the bounce $\omega < -1$ which turns out to be $0 > \lambda > -1/4$. The time instant when $\omega$ crosses the phantom divide line is calculated and reads $t \bigg |_{\omega = -1} = \pm \sqrt{\frac{2}{3\rho_{cr}}}$.
\item Next, expressions of density and pressure and a qualitative understanding of the initial conditions of the universe at the bounce ($t=0$) are derived. We found that the initial conditions (\textit{i.e,} density and pressure) of the universe are finite owing to the non-vanishing nature of the scale factor thus eliminating the initial singularity problem.
\item The violation of NEC and SEC at the bounce region is shown which is an important condition for achieving any non-singular bounce with standard matter sources (baryons, radiation, neutrinos, etc). 
\item Finally, the stability of the model is analyzed with respect to linear homogeneous perturbations in FLRW spacetime. Our model and hence matter bounce scenarios in general are highly unstable at the bounce in the framework of $f(R,T)$ gravity but the perturbations decay out rapidly away from the bounce safeguarding its stability at late times. 
\end{itemize} 

\section*{Acknowledgements} PS, SB, PKS acknowledges DST, New Delhi, India for
providing facilities through DST-FIST lab, Department of Mathematics, BITS-Pilani, Hyderabad campus where
a part of this work was done. One of the author (SB) thank Dr. Biswajit Pandey for stimulating discussions. We are very much grateful to the honorable referees and the editor for the illuminating suggestions that have significantly improved our work in terms of research quality and presentation.


\begin{thebibliography}{}
\bibitem{gr} A. Einstein, Annalen der Physik, \textbf{49} (1916) 769.
\bibitem {sr} A. Einstein, Annalen der Physik, \textbf{17} (1905) 891.
\bibitem {motion} M. Habibi, et al., ApJ, \textbf{847} (2017) 120.
\bibitem {eht} The Event Horizon Telescope Collaboration et al., ApJL, \textbf{875} (2019) L1.
\bibitem {ligo} B.P. Abbott, et al., Phys. Rev. Lett., \textbf{116} (2016) 061102.
\bibitem {prd} P. Pavlovic, M. Sossich, Phys. Rev. D, \textbf{95} (2017) 103519.
\bibitem{candidate} E. J. Copeland, M. Sami, S. Tsujikawa, Int. J. Mod. Phys. D, \textbf{15} (2006) 1753.
\bibitem{t} R. Femaro, F. Fiorini, Phys. Rev. D, \textbf{75} (2007) 084031.
\bibitem {r1} H. A. Buchdahl, Mon. Not. R. Astron. Soc., \textbf{150} (1970) 1.
\bibitem {r2} S. Nojiri, S. D. Odinstov, Phys. Rev. D, \textbf{68} (2003) 123512.
\bibitem {r3} T. P. Sotiriou, V. Faraoni, Rev. Mod. Phys., \textbf{82} (2010) 451.
\bibitem {r4} S. Capozziello, M. De Laurents, Phys. Rep., \textbf{509} (2011) 167.
\bibitem {r5} T. Clifton, et al., Phys. Rep., \textbf{513} (2012) 1.
\bibitem{harko/2011} T. Harko et al., Phys. Rev. D, {\bf 84} (2011) 024020.
\bibitem {g} S. Nojiri, et al., Phys. Suppl., \textbf{172} (2008) 81.
\bibitem{pk} P. K. Sahoo, S. K. Tripathy, P. Sahoo, Mod. Phys. Lett. A, \textbf{33} (2018) 1850193.
\bibitem{cmb} A. Penzias, R. Wilson,  Astrophys. J., \textbf{142} (1965) 419.
\bibitem{inflation} A. Guth, Phys. Rev. D, \textbf{23} (1981) 347.
\bibitem {instar} A. Starobinsky,  Phys. Lett. B, \textbf{91} (1980) 99.
\bibitem{flexible} R. Brandenberger, P. Peter, arXiv: 1603.05834v2 [hep-th].


%\bibitem{per1998} S. Perlmutter, et al., Astrophys. J., \textbf{517} (1998) 565.
%\bibitem{riess1998} A. Riess, et al., Astrophys. J., \textbf{116} (1998) 1009.
%\bibitem{bennet2003} C. L. Bennet, et al., Astrophys. J. Suppl. Ser., \textbf{148} (2003) 1.



\bibitem {c1} Yi-Fu Cai, T. Qiu, R. Brandenberger, X. Zhang, arXiv: 08103.4677v1[hep-th].
\bibitem {c2} Yi-Fu Cai, D. A. Easson,  R. Brandenberger, arXiv: 1206.2382v2[hep-th].
\bibitem {c3} Yi-Fu Cai, E. Wilson-Ewing, arXiv: 1412.2914(1)v2[gr-qc].
\bibitem {c4} Yi-Fu Cai, et al., arXiv: 1610.00938v2[astro-ph.CO]
\bibitem {c5}  Yi-Fu Cai, arXiv: 1405.1369v2[hep-th]
\bibitem {rb} R. Brandenberger, P. Peter, arXiv: 1603.05834v2[hep-th]
\bibitem {b1} K. Bamba, et al., J. Cosmol. Astropart. Phys. \textbf{1401} (2014) 008.
\bibitem {b2} K. Bamba, et al., Phys. Lett. B, \textbf{732} (2014) 349.
\bibitem {b3} K. Bamba, et al., J. Cosmol. Astropart. Phys., \textbf{1504}  (2015)  001.
\bibitem {b4} K. Bamba, et al., Phys. Rev. D ,\textbf{94}  (2016) 083513.
\bibitem {c6} A. de la Cruz-Dombriz, et al., Phys. Rev. D, \textbf{97}  (2018) 104040.
\bibitem {c7} Yi-Fu  Cai, et al., arXiv: 1104.4349v2[astro-ph.CO].

\bibitem{SKT19} S. K. Tripathy, R. K. Khuntia, P. Parida, Eur. Phys. J Plus, \textbf{134} (2019) 504.
\bibitem{h31} Y.-F. Cai, D. A. Easson,  R. Brandenberger, J. Cosmol. Astropart. Phys. \textbf{1208} (2012) 020 .
\bibitem{h19}  R. H. Brandenberger,  arXiv:1103.2271 [astro-ph.CO].
\bibitem{h19a} D. Battefeld, P. Peter, Phys. Rep., \textbf{571}  (2015)  1.
\bibitem{h19b} Y.-K. E. Cheung, X. Song,  S. Li, Y. Li, Y. Zhu, arXiv:1601.03807 [gr-qc].

\bibitem{h32} R. Brandenberger, arxiv: 1206.4196 [astro-ph.co]; R. Brandenberger,  Int. J. Mod. Phys. Conf. Ser. \textbf{01} (2011) 67; R. Brandenberger,  AIP Conf. Proc. \textbf{1268} (2010) 3.

\bibitem{h33} E. Wilson-Ewing, arxiv: 1211.6269 [gr-qc].

\bibitem{h34} Y. -F. Cai,  T. Qiu, Y. S. Piao,   M. Li, X. Zhang,  JHEP \textbf{0710} (2007) 071; Y.-F. Cai,  T. Qiu, R. Brandenberger, Y. S. Piao, X. Zhang,  J. Cosmol. Astropart. Phys. \textbf{0803} (2008), 013; Y. -F. Cai, X. Zhang,  J. Cosmol. Astropart. Phys. \textbf{0906} (2009) 003.

\bibitem{h37} T. Qiu, J. Evslin, Y. F. Cai, M. Li,  X. Zhang, J. Cosmol. Astropart. Phys. \textbf{1110} (2011), 036;  D. A. Easson,  I. Sawicki, A. Vikman, J. Cosmol. Astropart. Phys. \textbf{1111} (2011) 021.

\bibitem{h35} Y.-F. Cai,  T. Qiu,  R. Brandenberger, X. Zhang, Phys. Rev. D, \textbf{80} (2009) 023511.

\bibitem{h38} M. G. Brown, K. Freese,  W. H. Kinney, J. Cosmol. Astropart. Phys. \textbf{0803} (2008) 002; V. Dzhunushaliev,  V. Folomeev,   K. Myrzakulov, R. Myrzakulov, Int. J. Mod. Phys. D, \textbf{17} (2008) 2351; K. Nozari, S. D. Sadatian, Phys. Lett. B, \textbf{676} (2009) 1; E. N. Saridakis, S. V. Sushkov, Phys. Rev. D, \textbf{81} (2010) 083510; A. Banijamalia, B. Fazlpour, J. Cosmol. Astropart. Phys., \textbf{01} (2012) 039.

\bibitem{michael} M. S. Turner, AIP Conf. Proc., \textbf{478} (1999) 113.
\bibitem{h41} S. D. Odintsov,  V. K.  Oikonomou, Phys. Rev. D, \textbf{90} (2014) 124083.

\bibitem{h42}  S. D. Odintsov,  V. K. Oikonomou, Phys. Rev. D \textbf{92} (2015) 024016.
\bibitem{h44}  Y.-F. Cai, S.-H. Chen,  J. B. Dent, S. Dutta,  E. N. Saridakis, Class. Quantum Grav., \textbf{28} (2011) 215011;  K. Bamba,  G. G. L. Nashed,  W. El Hanafy, Sh. K. Ibraheem, Phys. Rev. D, \textbf{94} (2016) 083513.

\bibitem{h86}  Y. -F. Cai, T. Qiu,  R. Brandenberger, Y. S.  Piao, X. Zhang, J. Cosmol. Astropart. Phys., \textbf{03} (2015) 006.
\bibitem{sah60} M. Visser, C. Barcelo, COSMO \textbf{99} (1999) 98, arXiv:gr-qc/0001099. 
\bibitem{sah61} S. Carroll, Spacetime and Geometry: An Introduction to General Relativity (Addison Wesley, 2004).
\bibitem{sah65} M. Sharif, M. Zubair, J. Phys. Soc. of Japan, \textbf{82} (2013) 014002. 
%\bibitem{Odintsov/2014} S. D. Odintsov, V. K. Oikonomou, Phys. Rev. D, \textbf{90} (2014) 124083.

%\bibitem{Koehn/2014} M. Koehn, J. -L. Lehners, B. A. Ovrut, Phys. Rev. D, \textbf{90} (2014) 025005.
%\bibitem{odintsov/2015} S. D. Odintsov, V. K. Oikonomou, E. N. Saridakis, arXiv:1501.06591.
%\bibitem{Oikonomou/2014} V. K. Oikonomou, arXiv:1412.4343.

\end{thebibliography}
\end{document}